\documentclass[12pt,preprint]{aastex}
\usepackage{epsf}
\usepackage{natbib}
\usepackage{float}
\usepackage{appendix}
\usepackage{amsmath}
\usepackage{pdflscape}
\usepackage{gensymb}

\tabletypesize{\scriptsize}

\newcommand{\rom}[1]{\uppercase\expandafter{\romannumeral #1\relax}}
\newcommand{\appropto}{\mathrel{\vcenter{
  \offinterlineskip\halign{\hfil$##$\cr
    \propto\cr\noalign{\kern2pt}\sim\cr\noalign{\kern-2pt}}}}}

\def\gappeq{\mathrel{ \rlap{\raise.5ex\hbox{$>$}}
                      {\lower.5ex\hbox{$\sim$}}  } }

\begin{document}
\slugcomment{Accepted for Publication in ApJ}
\shorttitle{The Pseudo-zodi Problem}
\shortauthors{Stark, Kuchner, \& Lincowski}

\title{The Pseudo-zodi Problem for Edge-on Planetary Systems}

\author{Christopher C. Stark\altaffilmark{1}, Marc J. Kuchner\altaffilmark{2}, Andrew Lincowski\altaffilmark{3} }

\altaffiltext{1}{NASA Goddard Space Flight Center, Exoplanets \& Stellar Astrophysics Laboratory, Code 667, Greenbelt, MD 20771; christopher.c.stark@nasa.gov}
\altaffiltext{2}{NASA Goddard Space Flight Center, Greenbelt, MD 20771}
\altaffiltext{3}{The University of Arizona, Tucson, AZ 85721}

\begin{abstract}

Future direct observations of extrasolar Earth-sized planets in the habitable zone could be hampered by a worrisome source of noise, starlight-reflecting exozodiacal dust. Mid-infrared surveys are currently underway to constrain the amount of exozodiacal dust in the habitable zones around nearby stars.  However, at visible wavelengths another source of dust, invisible to these surveys, may dominate over exozodiacal dust.  For systems observed near edge-on, a cloud of dust with face-on optical depth $10^{-7}$ beyond $\sim5$ AU can mimic the surface brightness of a cloud of exozodiacal dust with equal optical depth if the dust grains are sufficiently forward-scattering.  We posit that dust migrating inward from cold debris belts via Poynting-Robertson drag could produce this ``pseudo-zodiacal" effect, potentially making it $\sim50\%$ as common as exozodiacal clouds.  We place constraints on the disk radii and scattering phase function required to produce the effect.
  
\end{abstract}


\section{Introduction}
\label{intro}

Our solar system hosts a diffuse cloud of dust near 1 AU generated by comets and asteroids called the zodiacal cloud.  By analogy, we expect extrasolar planetary systems to have similar ``exozodiacal" clouds in their habitable zones (HZ).  Mid-infrared observations have revealed exozodiacal clouds hundreds of times the brightness of the zodiacal cloud \citep[e.g.,][]{gaidos1999,bryden2006,hines2006,stark2009,smith2009,millangabet2011}, and we expect to find many more at fainter levels as detection limits improve.

Future NASA missions that aim to directly image extrasolar Earth-sized planets in the habitable zone must cope with starlight reprocessed by exozodiacal dust, which will act as a source of noise and substantially increase the required exposure time.  Most current mission concepts (e.g., WFIRST, Exo-C, Exo-S, and ATLAST) that could potentially detect extrasolar Earth-like planets would operate at visible wavelengths.  Thus, these missions would primarily deal with light \emph{scattered} by exozodiacal dust.

The most sensitive probes of dust in the HZ to date have come from interferometric observations in the mid-infrared, as blackbody radiation from dust in the HZ peaks at $\sim10$ $\mu$m.  Such observations have constrained the median exozodi level to $\lesssim 60$ ``zodis" \citep{mennesson2014}.  Future infrared observations with the Large Binocular Telescope Interferometer (LBTI) could achieve even greater sensitivity, potentially constraining the median exozodi level further \citep[e.g.,][]{kennedy2014}.

To apply these constraints to future direct-imaging missions, one must assume a correlation between the infrared excess attributable to exozodiacal dust and the visible wavelength surface brightness yet to be observed \citep{kennedy2014}.  Realistically, these two signals will not be exactly correlated, primarily due to the fact that micron-sized dust grains scatter visible wavelength light asymmetrically, predominantly in the forward direction.  As a result, visible wavelength observations of face-on exozodiacal dust clouds will appear significantly dimmer than expected from their albedo alone, while the discrepancy will be less for edge-on systems.  In addition, the appearance of edge-on systems will be complicated by the asymmetry of the scattering phase function---e.g., dust at larger circumstellar distances is observed at smaller scattering angles.  In fact, if sufficiently forward-scattering, dust exterior to the HZ could create a ``pseudo-zodiacal" flux that dominates over exozodiacal dust at visible wavelengths for near edge-on systems.

Naturally, many debris disks host massive clouds of dust exterior to the HZ.  Dust grains produced in these belts are affected by stellar radiation, including the relativistic Poynting-Robertson (PR) drag force, which removes angular momentum from the particles' orbits and causes them to migrate toward the host star over Myr time scales.  Dynamical studies have shown that in spite of the collisional destruction of dust grains, cold disks with optical depths $\sim 10^{-4}$ can deliver nearly 100 zodis\footnote{While many definitions of 1 ``zodi" exist, in this work we simply use the term to refer to an optical depth $\sim10^{-7}$.} of dust to the inner regions of the system via PR drag \citep[e.g.,][]{wyatt2005,stark2009,kuchner2010,vitense2010}.  Thus, dust just exterior to the HZ, and therefore the pseudo-zodiacal effect, may be common.

Here we discuss the impact of forward-scattering dust exterior to the HZ on the appearance of a debris disk at visible wavelengths as it applies to the problem of directly imaging Earth-like exoplanets.  In Section \ref{SPF_sec} we summarize estimates of the scattering phase functions of debris disks and other astrophysical dust sources.  In Sections \ref{pseudo_zodi_sec} and \ref{detection_sec} we introduce and quantify the impact of the pseudo-zodi effect, and show that the dust responsible for this effect is undetectable to any current telescope.  Finally, we discuss and summarize our findings in Section \ref{discussion_sec}.

\section{How forward-scattering is debris disk dust?}
\label{SPF_sec}

The scattering phase functions of debris disks in general are poorly understood.   Estimates of the degree of forward-scattering in observed debris disks often rely solely on the brightness ratio along the projected minor axis of the disk.  Typically these estimates fit the brightness ratio with a Henyey-Greenstein (HG) phase function of the form
\begin{equation}
	\Phi\left(\phi\right) = \frac{1}{4\pi}\frac{1-g^2}{[\,1+g^2-2g\cos{\phi}\,]^{3/2}},
\end{equation}
where $\phi$ is the scattering phase angle and $g = \int\nolimits\Phi\cos{\phi}\,\mathrm{d}\Omega$ is the scattering asymmetry parameter, ranging from $-1$ for perfect back-scattering to $1$ for perfect forward-scattering. Fits to disk observations using this method typically arrive at $g$ values ranging from $0.0$ to $0.3$ \citep[e.g.][]{kalas2005,schneider2006,debes2008,thalmann2011}, significantly less forward-scattering than the $\int\nolimits\Phi\cos{\phi}\,\mathrm{d}\Omega\sim0.9$ values commonly predicted by Mie theory \citep[e.g.,][]{rodigas2014}.

However, we should not necessarily expect HG fits to correctly determine the true degree of forward scattering in debris disks.  First, the HG SPF is a mathematical construct designed such that $g = \int\nolimits\Phi\cos{\phi}\,\mathrm{d}\Omega$, the first moment of the scattering phase function, and is not a physical model.  HG SPFs cannot necessarily reproduce the \emph{shape} of  observed scattering phase functions \citep[e.g.,][]{stark2014}.

Second, we cannot observe debris disks over all scattering angles---we are limited by the disk inclination that nature provides to scattering angles $\pi/2 - i < \phi < \pi/2 + i$.  The smallest and largest scattering angles are also commonly unobservable due to the inner working angle (IWA) of high contrast images.  This limited range of scattering angles prevents observations of the forward-scattering peak, where the majority of the change in an SPF occurs.  As a result, fits to a disk's brightness variations near $\phi=\pi/2$ are highly degenerate.  Furthermore, images of edge-on systems, which have the largest range of observable scattering angles, suffer from degeneracies between the SPF and radial dust distribution.  

Recent observations of debris disks suggest that the SPF may be more forward-scattering than previously thought.
Variations in the brightness of the HD 181327 debris disk suggest a non-HG SPF with a strong increase at the smallest observable scattering angles $\sim60\degree$ \citep{stark2014}.  Observations of the HR 4796A debris disk reveal a potentially complex SPF, possibly requiring strongly forward-scattering grains in an optically thick disk \citep{perrin2014}.  Very forward-scattering dust would also help explain the low apparent albedos of some debris disks \citep{stark2014}.

Other astrophysical dust sources appear to exhibit strong forward-scattering as well.  \citet{hong1985} provided estimates of the zodiacal cloud's scattering phase function, showing moderately strong forward-scattering ($g\sim0.7$), with a relatively flat, non-HG profile near $\phi=\pi/2$.  Unfortunately these estimates were limited to scattering angles $\gtrsim30\degree$.  Recent observations of the zodiacal cloud can achieve much smaller scattering angles than observed extrasolar debris disks, e.g. STEREO can observe at solar elongations of a few degrees.  However, estimates of the zodiacal cloud's SPF suffer from degeneracies between the SPF, the radial dust density distribution, and the dependence of the size distribution on circumstellar distance.

Observations of the ISM dust have also suggested strongly forward-scattering grains.  \citet{gibson2003} showed that forward-scattering of a nearby dust cloud with $g=0.74$ best explained a UV excess in the Pleiades reflection nebula.  Similarly, \citet{murthy2011} discovered near-UV halos around several nearby, bright stars, which they attribute to optically thin nearby ISM dust clouds with $g=0.72$.

\section{The ``pseudo-zodi" effect}
\label{pseudo_zodi_sec}

\begin{figure}[H]
\begin{center}
\includegraphics[width=6.5in]{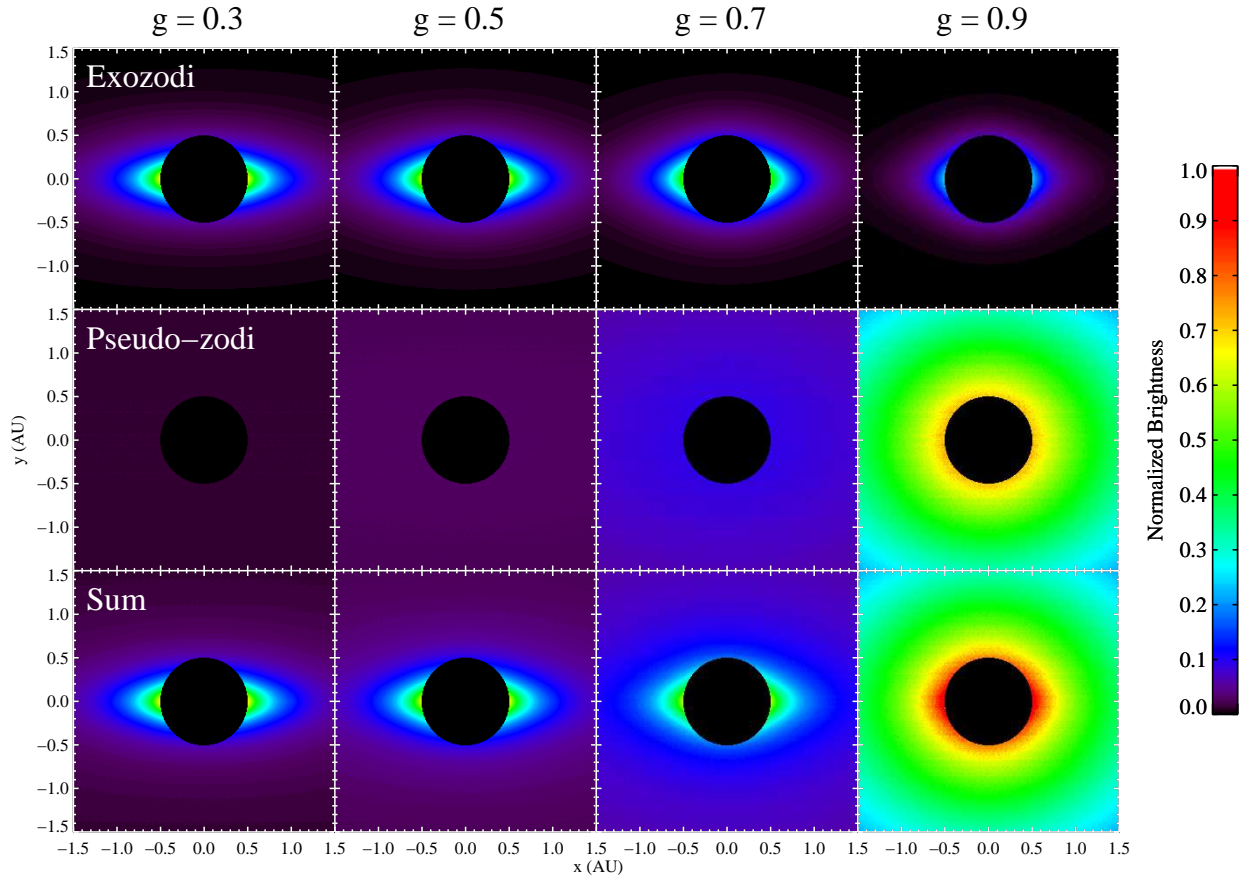}
\caption{Synthetic visible-wavelength images of exozodi and pseudo-zodi, based on our solar system's debris disk, viewed edge-on as a function of scattering asymmetry parameter $g$ assuming a HG SPF.  The central $0.5$ AU has been masked off.  At a projected separation of 1 AU, the pseudo-zodi becomes brighter than the exozodi for $g>0.7$. \label{ss_edgeon}}
\end{center}
\end{figure}

To illustrate how dust exterior to the HZ affects the visible wavelength appearance of an edge-on disk, we produced synthetic images of the solar system's debris disk as it would appear edge-on from afar, as shown in Figure \ref{ss_edgeon}.  For the ``exozodiacal" component, we synthesized images of the zodiacal cloud using ZODIPIC \citep{moran2004}.  We modified ZODIPIC to use a HG SPF and adopted an outer edge of 3 AU.  For the ``pseudo-zodiacal" component, we synthesized images of the Kuiper Belt dust with \emph{dustmap} \citep{stark2011} using the dynamical model of \citet{kuchner2010} with optical depth $\tau\sim10^{-7}$.  This model, like other Kuiper Belt dust models, predicts dust extending inward of the Kuiper Belt to $\sim5$ AU due to Poynting-Robertson drag.  Jupiter ejects inward-migrating KB dust, such that the model KB dust density is negligible compared to the zodiacal cloud interior to 5 AU.  For this investigation, we explicitly removed all KB dust interior to 5 AU, where the KB dust models become less reliable due to Poisson noise.  To further reduce Poisson noise, we azimuthally averaged the Kuiper Belt dust model around the disk's axis of symmetry in steps of $0.36\degree$.  For Figure \ref{ss_edgeon}, we masked off the central $0.5$ AU of each image.

As the SPF becomes more forward-scattering, the brightness of the pseudo-zodiacal component increases because dust at larger distances is observed at smaller scattering angles.  We measured the surface brightness of the model disks at a projected separation of 1 AU.  We find that the brightness of the pseudo-zodi (Kuiper Belt dust) exceeds that of the exozodi (zodiacal dust) for $g > 0.7$.    Equivalently, \emph{in the complete absence of any true exozodiacal light, an edge-on Kuiper Belt dust cloud analog could produce a brightness equal to 1 zodi of exozodiacal light with $g \gtrsim 0.7$}.

\begin{figure}[H]
\begin{center}
\includegraphics[width=6.5in]{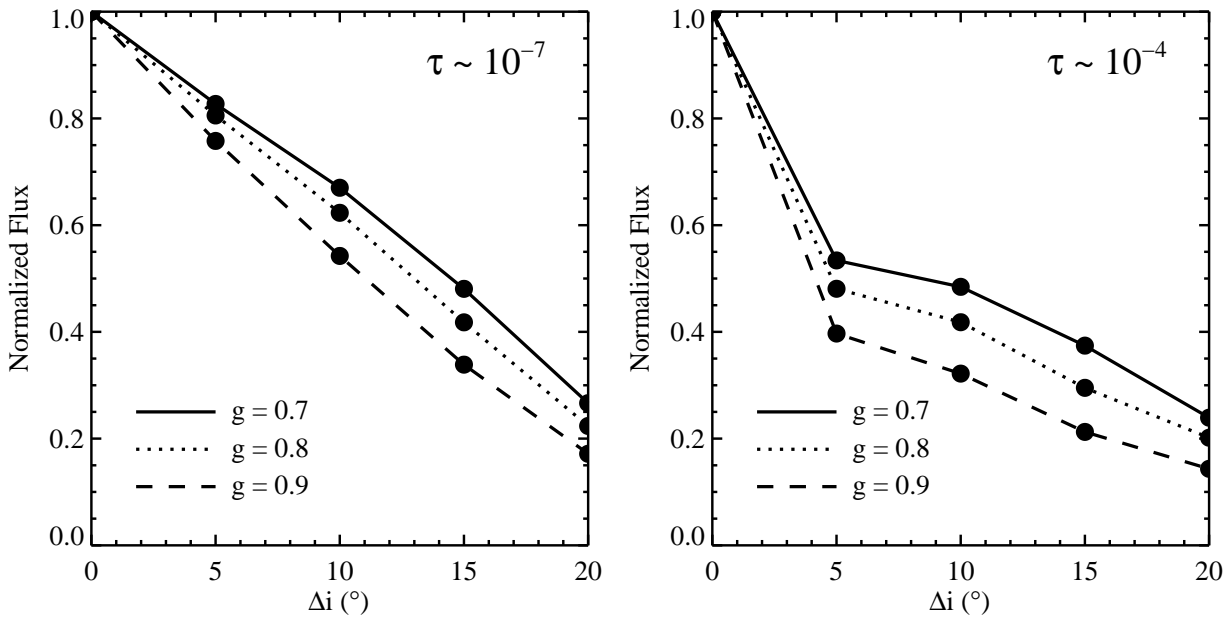}
\caption{Normalized pseudo-zodiacal surface brightness at a projected separation of 1 AU along the disk's projected major axis as a function of inclination (in degrees from edge-on) and $g$.  The pseudo-zodi effect is significant over a range of inclinations consistent with the opening angle of the disk. \label{kb_vs_i}}
\end{center}
\end{figure}

The pseudo-zodi effect persists over a range of inclinations roughly equal to the opening angle of the disk $H/r$, where the scale height $H$ is effectively the HWHM of the disk's vertical density distribution, and $r$ is circumstellar distance.  Using the KB dust models of \citet{kuchner2010}, we produced images like those shown in Figure \ref{ss_edgeon} at different inclinations and disk optical depths.  Figure \ref{kb_vs_i} shows how the pseudo-zodiacal surface brightness decreases as the disk deviates from an edge-on orientation for the $\tau\sim10^{-7}$ and $\tau\sim10^{-4}$ disk models.  The opening angle of the $\tau\sim10^{-7}$ KB dust model is $\sim13\degree$, roughly equal to the inclination at which the pseudo-zodi signal is reduced by a factor of 2.  The opening angle of the $\tau\sim10^{-4}$ disk is substantially smaller becase grain-grain collisions tend to remove particles on inclined orbits.

Now that we have considered the specific case of dust in the solar system, let us consider a more general dust cloud model to understand the conditions required to create a substantial pseudo-zodi effect. In particular, we will calculate what the inner edge of the outer disk must be to produce a surface brightness equivalent to that of the habitable zone dust.  We adopt a simpler analytic model composed of two concentric circular debris belts.  We assume radially-uniform vertical optical depths, appropriate for Poynting-Robertson drag-dominated disks, of $\tau_{\rm ID}$ and $\tau_{\rm OD}$ for the inner and outer disk, respectively.  For simplicity, we assume both disks have equal opening angles, $\sigma_i$, with scale heights given by $H=r \tan{\sigma_i}$, where $r$ is circumstellar distance.  We assume the inner edge of the inner disk is small compared to the observational inner working angle, and adjust only its outer edge, $r_{\rm ID,out}$.  The outer disk has an inner edge $r_{\rm OD,in}$ and extends to infinity, a valid approximation since dust near the inner edge dominates the brightness integral.

For an edge-on system, the brightness integral along the line of sight in the mid-plane at a projected separation of $s$ is given by
\begin{equation}
	B\!\left(s\right) \appropto \int \! \frac{\tau}{2H\!\left(r\right)}\; r^{-2}\; \Phi\!\left(\phi\right) \mathrm{d}z,
\end{equation}
where $z$ measures line-of-sight distance with zero corresponding to the distance of the star, $\Phi\!(\phi)$ is the scattering phase function, $\phi = \tan^{-1}{(s/z)}$ is the scattering angle, and the integral is to be taken over the appropriate limits on $z$.  Using the above equation, we can express the brightness of the inner disk as
\begin{equation}
	B_{\rm ID}\!\left(s\right) \appropto \frac{\tau_{\rm ID}}{2\tan\sigma_i}\; \int_{-z_{\rm ID,out}}^{z_{\rm ID,out}} \! r^{-3}\; \Phi\!\left(\phi\right) \mathrm{d}z,
\end{equation}
where $z_{\rm ID,out} = (r_{\rm ID,out}^2 - s^2)^{1/2}$.  For the outer disk,
\begin{equation}
	B_{\rm OD}\!\left(s\right) \appropto \frac{\tau_{\rm OD}}{2\tan\sigma_i}\; \int_{-\infty}^{-z_{\rm OD,in}} \! r^{-3}\; \Phi\!\left(\phi\right) \mathrm{d}z + \frac{\tau_{\rm OD}}{2\tan\sigma_i}\; \int_{z_{\rm OD,in}}^{\infty} \! r^{-3}\; \Phi\!\left(\phi\right) \mathrm{d}z,
\end{equation}
where $z_{\rm OD,in} = (r_{\rm OD,in}^2 - s^2)^{1/2}$.
We can express the brightness ratio of the outer and inner disks as
\begin{equation}
	\label{bratio}
	\frac{B_{\rm OD}\!\left(s\right)}{B_{\rm ID}\!\left(s\right)} = \frac{\tau_{\rm OD}}{\tau_{\rm ID}}\frac{\int_{-\infty}^{-z_{\rm OD,in}} \! r^{-3}\; \Phi\!\left(\phi\right) \mathrm{d}z + \int_{z_{\rm OD,in}}^{\infty} \! r^{-3}\; \Phi\!\left(\phi\right) \mathrm{d}z}{\int_{-z_{\rm ID,out}}^{z_{\rm ID,out}} \! r^{-3}\; \Phi\!\left(\phi\right) \mathrm{d}z}.
\end{equation}

Equation \ref{bratio} shows that the ratio of surface brightnesses is the product of the optical depth ratio and a geometric factor, which we can calculate by numerically evaluating three integrals.  Given an outer edge for the inner disk $r_{\rm ID,out}$, an optical depth ratio $\tau_{\rm OD}/\tau_{\rm ID}$, and a scattering phase function $\Phi(\phi)$, we can set $B_{\rm OD}(s)/B_{\rm ID}(s)=1$ and solve for the maximum $r_{\rm OD,in}$ for which the outer disk produces the same brightness as the inner disk.

\begin{figure}[H]
\begin{center}
\includegraphics[width=4in]{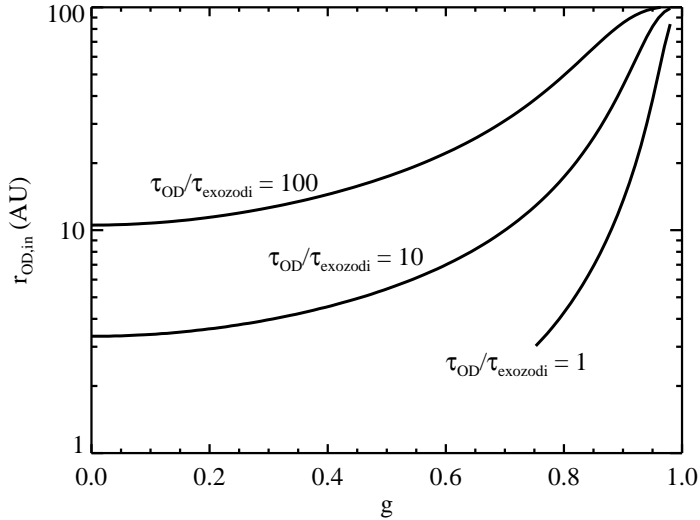}
\caption{Inner radius and forward scattering parameter $g$ at which the surface brightness of an outer dust belt equals the the surface brightness of exozodiacal dust at a projected separation of 1 AU in an edge-on system.  If $g=0.75$, 1 zodi of dust beyond 3 AU can mimic 1 zodi of exozodiacal dust.\label{bratio_fig}}
\end{center}
\end{figure}

Assuming $r_{\rm ID,out} = 3$ AU (roughly appropriate for an exozodiacal cloud) and assuming the same HG SPF for both inner and outer disks, we calculated what the inner edge of the outer disk must be to produce the exozodiacal-equivalent brightness as a function of $g$.  Figure \ref{bratio_fig} shows this limit for different ratios of $\tau_{\rm OD}/\tau_{\rm ID}$.  For $\tau_{\rm OD}/\tau_{\rm ID} = 1$, dust exterior to 3 AU begins to dominate the brightness integral for $g>0.75$, i.e. one zodi of dust exterior to 3 AU could mimic one zodi of exozodiacal dust if $g>0.75$.  Equivalently, the $\tau_{\rm OD}/\tau_{\rm ID} = 1$ line also shows that 10 zodis of dust exterior to 3 AU could mimic 10 zodis of exozodiacal dust if $g>0.75$.  If $g=0.85$, the outer disk could be as distant as 7 AU.  As the $\tau_{\rm OD}/\tau_{\rm ID} = 10$ line shows, 10 zodis of dust exterior to $\sim3$ AU can mimic one zodi of exozodiacal dust, even in the case of isotropic scattering.

\begin{figure}[H]
\begin{center}
\includegraphics[width=4in]{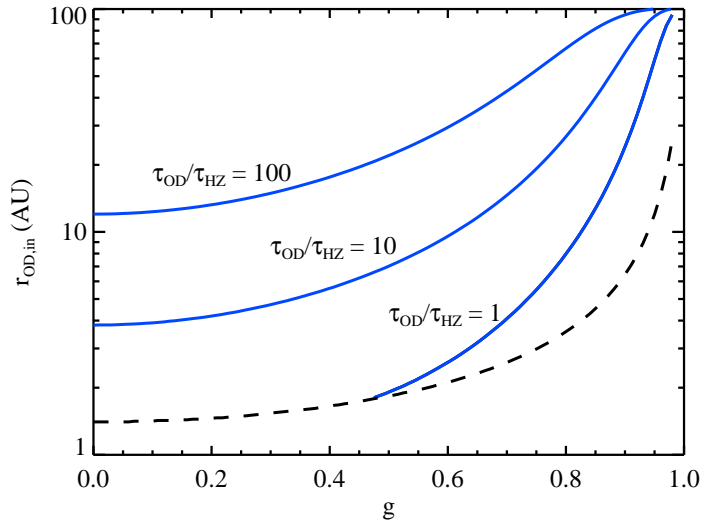}
\caption{Inner radius and forward scattering parameter $g$ at which the surface brightness of an outer dust belt equals the surface brightness of habitable zone dust (truncated at $1.77$ AU) at a projected separation of 1 AU in an edge-on system (blue curves).  The dashed line shows the radius at which the outer region of a single, continuous debris disk dominates over the inner region.\label{bratio_fig2}}
\end{center}
\end{figure}

We can make similar curves using alternative definitions for the outer edge of the inner disk.  For example, the blue curves in Figure \ref{bratio_fig2} correspond to $r_{\rm out} = 1.77$ AU, the outer edge of the classical habitable zone \citep{kopparapu2013}.  Another useful limit that is independent of the HZ or exozodi definition comes from setting $r_{\rm out} = r_{\rm in}$, i.e. a single continuous disk.  The black dashed line in Figure \ref{bratio_fig2} plots the circumstellar distance beyond which the outer region of an edge-on uniform disk dominates the brightness integral.

\section{The invisibility of pseudo-zodi dust to current telescopes}
\label{detection_sec}

\subsection{LBTI}
Could we detect the dust contributing to the pseudo-zodi effect at N band with LBTI?  To answer this, we used \emph{dustmap} to create synthetic 10 $\mu$m images of the \emph{most massive} KB dust model ($\tau\sim10^{-4}$) generated by \citet{kuchner2010}.  This model produces PR-drag migrated dust interior to 35 AU within a factor of 2 of the theoretical $\tau = v_{\rm kep}/c$ limit \citep{kuchner2010}.  The \emph{dustmap} code calculates scattered light and thermal emission using Mie theory and self-consistently calculates grain temperature by balancing input and output energy.  We adopted optical constants appropriate for astronomical silicates \citep{li2001}.  

We then calculated the null depth of the most massive KB dust model by multiplying the model image by the transmission map of LBTI as described in \citet{kennedy2014}, and taking the ratio of the transmitted flux to the total disk plus stellar flux.  We aligned the interferometric fringes perpendicular to the disk midplane to maximize the null depth and produced images 40 AU $\times$ 40 AU to capture nearly all leaked flux.  The most massive edge-on KB dust model produced a null depth of $8\times10^{-5}$, smaller than the estimated LBTI null depth uncertainty of $10^{-4}$ \citep{kennedy2014}.  Thus, we conclude that the dust responsible for the pseudo-zodiacal effect will be invisible to LBTI.

\subsection{VLTI}

Could we detect the pseudo-zodi effect with VLTI?  The dust responsible for the pseudo-zodiacal light may be more detectable at shorter wavelengths, where we directly measure the forward scattering pseudo-zodi effect.  A number of NIR excesses have been detected interferometrically at H and K band around nearby stars \citep[e.g.,][]{absil2006,difolco2007,absil2008,akeson2009,absil2013}, potentially suggesting the presence of large amounts of thermally emitting hot dust.  However, seven out of nine NIR excesses detected with VLTI/PIONIER by \citet{ertel2014} show wavelength-independent flux ratios, indicative of scattered light.  \citet{ertel2014} ruled out the possibility of forward-scattering, arguing that no cold dust has been detected around such systems, but did not address the possibility of dust at intermediate distances.  \citet{defrere2012} showed that as much as 70\% of the NIR flux observed in the edge-on $\beta$ Pic system could come from forward-scattering of starlight by dust beyond 10 AU.  \citet{defrere2012} concluded that the remaining 30\% therefore demanded hot dust, though they did not include the contribution from the inclined sub-disk or consider dust at distances $\sim5$ AU.

\begin{figure}[H]
\begin{center}
\includegraphics[width=4.5in]{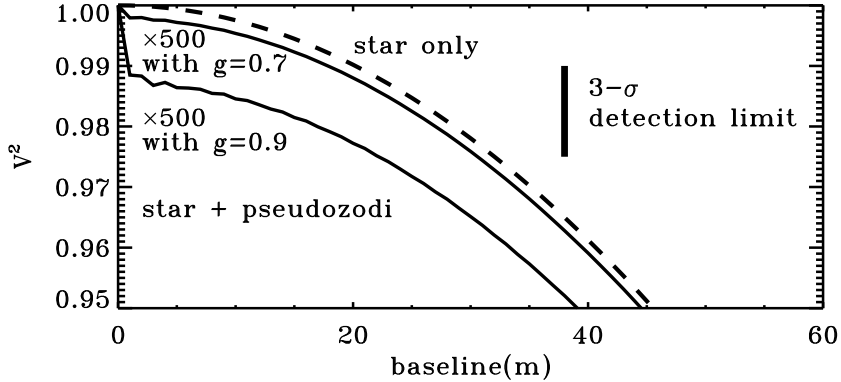}
\caption{Visibility amplitude as a function of baseline at H band for a Solar twin with and without the $\tau\sim10^{-4}$ KB dust model of \citet{kuchner2010} (solid and dashed lines, respectively).  The disk's surface brightness was increased by a factor of 500 for illustration.  The $3\sigma$ detection threshold used by \citet{ertel2014} shows that dust contributing to pseudo-zodi would go undetected by VLTI.\label{vlti_fig}}
\end{center}
\end{figure}

Could the pseudo-zodi effect explain these NIR detections?  Figure \ref{vlti_fig} shows the estimated H-band visibility as a function of baseline for the most massive ($\tau\sim10^{-4}$) \citet{kuchner2010} simulation, for comparison with Figure 1 from \citet{ertel2014}.  Although the $\tau\sim10^{-4}$ model is within a factor of 2 of the theoretical $\tau\sim v_{\rm kep}/c$ limit, we have artificially increased the disk's surface brightness by a factor of 500 to illustrate the visibility deficit.  Given the $3\sigma$ detection threshold implemented by \citet{ertel2014}, these models cannot explain the NIR interferometric detections.

\subsection{\emph{Spitzer}}

\begin{figure}[H]
\begin{center}
\includegraphics[width=6.5in]{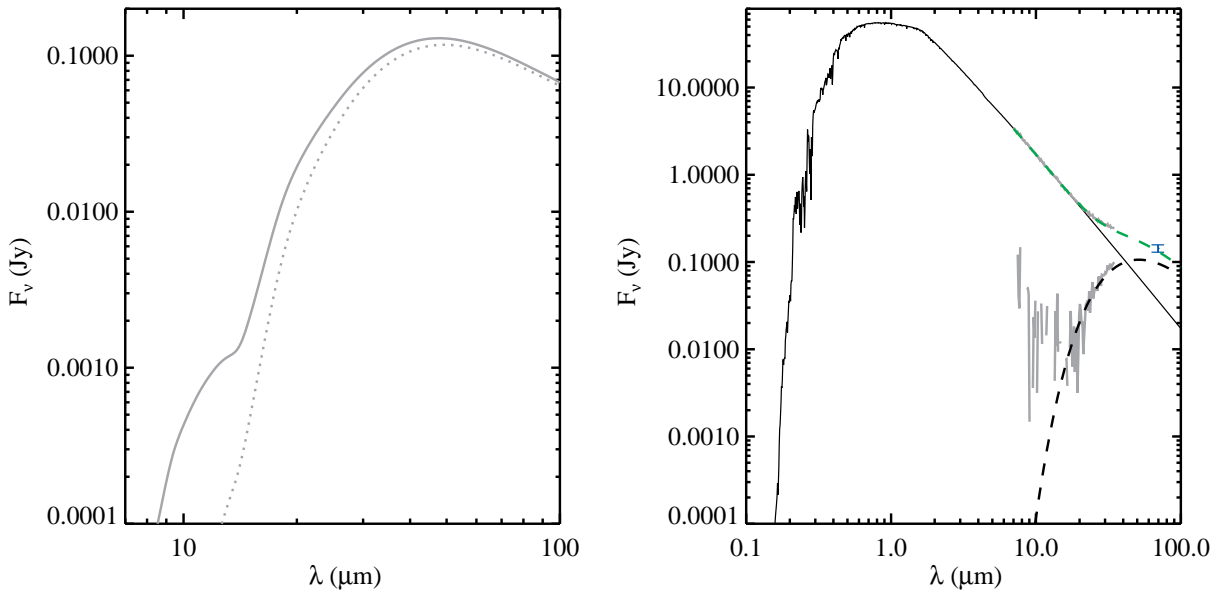}
\caption{\emph{Left:} SED for the $\tau\sim10^{-4}$ \citet{kuchner2010} KB dust model with and without PR-migrated dust interior to 35 AU (solid and dotted lines, respectively).  The dust interior to 35 AU that produces pseudo-zodi has a small affect on the SED. \emph{Right:} Simulated observations of the Sun and KB dust model with dust interior to 35 AU.  Simulated IRS data with and without the stellar photosphere subtracted are shown in gray, simulated MIPS 70 $\mu$m data is shown in blue, and the best fit single-component blackbody with and without the stellar photosphere subtracted are shown as dashed lines (compare to Figure 4 in \citet{chen2014}).  The dust contributing to pseudo-zodi would go undetected.\label{sed_fig}}
\end{center}
\end{figure}

Could we detect the dust contributing to the pseudo-zodi effect by looking for an IR excess in the star's spectral energy distribution (SED)?  To answer this, we used \emph{dustmap} to create synthetic SEDs of the $\tau\sim10^{-4}$ KB dust model.  The left panel in Figure \ref{sed_fig} shows the calculated SED for the $\tau\sim10^{-4}$ KB dust disk model as a solid line.  The dotted line shows the SED of the same model, but with dust interior to 35 AU removed.  PR drag-migrated dust, which we posit as a likely cause of the potential pseudo-zodi effect, marginally broadens the profile of the SED and creates a small bump at $\lambda\sim10$ $\mu$m.

To determine the detectability of the small bump in the SED caused by PR drag-migrated dust, we simulated \emph{Spitzer} Infrared Spectrograph (IRS) and Multiband Imaging Photometer (MIPS) observations for this disk around a Sun-like star at 10 pc.  The right panel of Figure \ref{sed_fig} shows the synthetic IRS (gray) and MIPS 70 $\mu$m data (blue) for the KB dust model shown as a solid line in the left panel, along with the assumed Kurucz stellar atmosphere model for the Sun \citep{castelli2004}.  We added normally distributed noise to the synthetic data, taken to be 2\% and 10\% of the total flux for IRS and MIPS, respectively, somewhat smaller than the uncertainties of real data sets \citep{chen2014}.  We note that the apparent excess at $\lambda<15$ $\mu$m is due to the noise of the synthetic IRS spectrum, not the SED of the model.

We then analyzed the synthesized data using methods similar to those of \citet{chen2014}.  We fit the synthetic data with one- and two-component blackbody models, sampling temperatures from 30 to 500 K in steps of $1.2$ degrees K.  The best single component blackbody model selects a temperature of $99.4$ K, with reduced $\chi^2/\nu = 0.79$.  The best two-component model selects two nearly identical belts separated by 1 degree K and provides only a 3\% reduction in the reduced $\chi^2/\nu$ statistic compared to the single-component model; a two-component blackbody model does not improve the fit and does not provide any information about the PR drag-migrated dust.  We conclude that PR-drag migrated dust responsible for the pseudo-zodi effect would largely go undetected by SED analyses.

\section{Discussion \& Conclusions}
\label{discussion_sec}

We have shown that dust exterior to the HZ may dominate the visible wavelength flux at a projected separation of 1 AU in systems close to edge-on.  This pseudo-zodiacal light, potentially much brighter than 1 zodi, can be created by disks with optical depths as little as $\sim10^{-7}$, but requires forward-scattering dust ($g\gtrsim0.7$ assuming a HG SPF).  This dust would likely not be detectable using current observatories.

Our brightness integral analysis above assumed a uniform optical depth for the inner and outer disks.  Observations show that the zodiacal cloud's optical depth may slightly decrease with distance from the Sun \citep{kelsall1998}.  Dynamical models suggest that the drag-dominated Kuiper Belt dust cloud's optical depth may increase with circumstellar distance in the region of 10--40 AU \citep{kuchner2010}.  Both of these trends may serve to enhance the pseudo-zodiacal light.  We also assumed edge-on disks with equal opening angles, $\sigma_i$, and equal scattering phase functions.  In reality, if the opening angles of the inner and outer disks are different or the degree of forward scattering is different, the amount of pseudo-zodiacal light will be impacted.  

We have no compelling estimates of the fraction of stars that host dust near $\sim5$ AU with optical depths $\gtrsim10^{-7}$.  However, because the pseudo-zodi effect persists over a range of inclinations roughly equal to the opening angle of the disk, and the probability of viewing a planetary system near edge-on is greater than near pole-on, pseudo-zodiacal light could be commonplace.  The opening angle of the hot component of the Kuiper Belt is $\sim15\degree$ \citep{trujillo2000}.  If all faint debris disks have similar opening angles, roughly 25\% of stars would be oriented properly to produce the pseudo-zodi effect.  Modeling of $\sim500$ debris disks observed with the \emph{Spitzer} IRS and MIPS at 70 $\mu$m \citep{chen2014} suggests that cold (80--180 K) debris disks are roughly twice as common as hot ($\sim340$ K) ones. Extrapolating this result to Kuiper Belt analogs and assuming an isotropic distribution of viewing angles would imply that the pseudo-zodiacal effect should be roughly 50\% as common as habitable zone zodiacal clouds.

We may also observe a similar, though likely weaker, effect in systems that are not viewed edge-on.  Numerous cold debris disks show warped, extended halos consistent with dust blown by the ISM \citep[e.g.][]{hines2007, debes2009}.  If the geometry of these dust halos are such that dust passes in front of the star, an ISM-blown dust veil could produce the same pseudo-zodiacal light.

The degree of forward scattering required to produce the pseudo-zodi effect could also produce other exotic disk phenomena.  As discussed in \citet{stark2011}, highly forward scattering dust can lead to periodic variations in the disk flux due to the orbital motion of clumpy dust structures; as a clump passes in front of the star, the disk flux increases.  These clumpy structures may appear to ``blink" on and off as they orbit their host star and pass through regions of small scattering angles, even in systems not aligned perfectly edge-on.

Further constraints on the scattering phase function of debris dust are necessary to determine the possibility of the pseudo-zodi effect.  However, the apparently high degree of forward-scattering for zodiacal cloud and ISM dust make this reasonably plausible.  Future missions that aim to directly image Earth-like planets may have to adjust their observation plans for edge-on systems because of this effect.

\acknowledgments

This research was supported by an appointment to the NASA Postdoctoral Program at Goddard Space Flight Center, administered by Oak Ridge Associated Universities through a contract with NASA.

\bibliography{*.bbl}

\end{document}